\documentclass[11pt,oneside]{article}

\usepackage[numbers,compress]{natbib}
\setcitestyle{numbers}

\usepackage{ragged2e}
\usepackage{placeins}
\usepackage{pbox}

\usepackage{amsmath, amssymb, amsthm, amsfonts, graphicx, mathrsfs, array, stackrel, enumerate}
\usepackage{url, overpic, times, eurosym}
\usepackage{booktabs, multirow}
\usepackage{bigdelim}
\usepackage{caption,newfloat}
\DeclareCaptionType{InfoBox}

\usepackage{pifont}
\usepackage{bm}
\usepackage{wrapfig}

\usepackage[table]{xcolor}
\definecolor{lemoncolor}{RGB}{255, 245, 160} 

\newcommand{\matrva}[1]{\bm{#1}} 

\newcommand{\bmu}{\matrva{\mu}}
\newcommand{\bSigma}{\matrva{\Sigma}}
\newcommand{\bPsi}{\matrva{\Psi}}
\newcommand{\bOmega}{\matrva{\Omega}}
\newcommand{\btheta}{\matrva{\theta}}
\newcommand{\beps}{\matrva{\epsilon}}
\newcommand{\bbeta}{\matrva{\beta}}

\newcommand{\bA}{\matrva{A}}
\newcommand{\bB}{\matrva{B}}
\newcommand{\bP}{\matrva{P}}
\newcommand{\bI}{\matrva{I}}

\newcommand{\bG}{\matrva{G}}
\newcommand{\bg}{\matrva{g}}
\newcommand{\bef}{\matrva{f}}
\newcommand{\bF}{\matrva{F}}
\newcommand{\bx}{\matrva{x}}
\newcommand{\by}{\matrva{y}}
\newcommand{\br}{\matrva{r}}

\newcommand{\bw}{\matrva{w}}

\newcommand{\bX}{\matrva{X}}
\newcommand{\bz}{\matrva{0}}
\newcommand{\bo}{\matrva{1}}
\newcommand{\bD}{\matrva{D}}

\usepackage{tabularx}
\newcolumntype{Y}{>{\centering\arraybackslash}X}

\usepackage{float}
\restylefloat{table}



\allowdisplaybreaks

\title{On statistical arbitrage under a conditional factor model of equity returns.}
\author{Trent Spears\thanks{University of Oxford, Oxford-Man Institute of Quantitative
    Finance.} \thanks{
    Corresponding author.  E-mail: trent@robots.ox.ac.uk}, Stefan Zohren, Stephen Roberts
}
\date{\today}

\begin{document}
\thispagestyle{plain} \maketitle

\renewcommand{\abstractname}{Summary}
\begin{abstract}
\noindent
We consider a conditional factor model for a multivariate portfolio of United States equities in the context of analysing a statistical arbitrage trading strategy.  A state space framework underlies the factor model whereby asset returns are assumed to be a noisy observation of a linear combination of factor values and latent factor risk premia.  Filter and state prediction estimates for the risk premia are retrieved in an online way.  Such estimates induce filtered asset returns that can be compared to measurement observations, with large deviations representing candidate mean reversion trades.  Further, in that the risk premia are modelled as time-varying quantities, non-stationarity in returns is de facto captured.  We study an empirical trading strategy respectful of transaction costs, and demonstrate performance over a long history of 29 years, for both a linear and a non-linear state space model.  Our results show that the model is competitive relative to the results of other methods, including simple benchmarks and other cutting-edge approaches as published in the literature. Also of note, while strategy performance degradation is noticed through time -- especially for the most recent years -- the strategy continues to offer compelling economics, and has scope for further advancement.  \\

\end{abstract}

\section{Introduction}

Statistical arbitrageurs seek trading opportunities whereby an asset price (or return) is dislocated from a measure of fair value to which it is assumed it will revert.  Such dislocation needs to be sufficiently large so as to be profitably traded -- at least, in expectation.  These strategies are documented as originating within the financial services in the late 1970s and early 1980s, within both the hedge fund and investment banking industries \cite{Thorp03}.  To this day, statistical arbitrage strategies are of popular practical application, and strategy techniques and insights have inspired a deep academic literature.  A comprehensive review, for work dated up till 2016, can be found in \cite{Kra17}.

\vspace{0.25 \baselineskip}

\noindent Indeed, much of the academic literature on statistical arbitrage ultimately concerns estimating a fair value model for an asset spread or portfolio, and/or modelling and analysing the stochastic dynamics of the dislocation about fair value.  This paper is mostly concerned with the former.  To our knowledge, we are the first authors to present and assess a conditional factor model for estimating fair value in the context of statistical arbitrage.  The model assumes that future returns are predictable given (i) the asset-wise factor exposures for the universe of assets of interest, and (ii) a latent factor vector that is estimable by regression in the asset cross-section.  Further, we assume the dynamic nature of the factor vector, suggesting a state space framework for the joint modelling of factors and asset returns.

\vspace{0.25 \baselineskip}

\noindent Our fair value model is closest in spirit to existing work that presents explanatory models of returns based on Principal Component Analysis (PCA) \cite{Ave10, Foc16}.  In these papers, the principal components (and their loadings) that are deemed to have sufficiently high explanatory power are an analogue to our model's factor exposures (and factor vectors).  However the papers differ significantly in many key workflow assumptions relative to ours, most importantly with respect to data set construction, asset feature modelling, and the trading strategy assumed.  We posit, critically, that the details of each of these steps can have a significant impact on the estimated performance statistics.  Hence we clearly describe our methods in the interest of ease of interpretability and reproducibility, and present a range of performance data based on competing sets of reasonable assumptions.  Owing to transparency, we also find two key, well-known works with which we can directly compare our results; this is less commonly found possible for academic quantitative finance papers that depend on strategy backtests.  In any case, we further describe these comparable papers, and other recent work in the spirit of ours, in Section 2 below.  This summary mostly covers work not contained in the aforementioned review \cite{Kra17}, due to their recency.

\vspace{0.25 \baselineskip}

 \noindent In Section 3, we present further the state space framework as it relates to the conditional factor model.  We model returns as a noisy observation of the latent factor vectors, and describe the well-known Kalman Filter algorithm for the case of a linear model, and the Unscented Kalman Filter algorithm for the case the model specified is non-linear.  In Section 4 we review details of the dataset construction, and the factor dynamics model.  We also discuss the particulars of the trading strategy, that amounts to the management of a long-short beta neutral equity portfolio.  We present the specification of the trading rule, realistic transaction cost assumptions, the hedging of market risk, and the calculation of strategy returns and bankroll management.  

\vspace{0.25 \baselineskip}

 \noindent Our results are presented in Section 5.  We show that the conditional factor state space model is useful for trading mid-frequency statistical arbitrage in US equity markets.  This is true with respect to a meaningful benchmark model and a simpler ordinary least squares-based approach, and to the results given in other well-known academic studies calculated over similar data sets.  We conclude in Section 6, and given that the model presents as a compelling starting point for building a strategy at larger scale, we include notes for both academic and practitioner future work.

\section{Existing literature}

We collect and describe studies relating to statistical arbitrage trading, for work mostly post the review period of \cite{Kra17}.   Unsurprisingly, many recent methods for approaching the problem subsume modern machine learning techniques.  These approaches mostly concern pairs trading.  For example, surrendering performance interpretability, the work of \cite{Sar20} presents a spectrum of techniques -- Principal Component Analysis (PCA) for dimension reduction of raw data, clustering as a strategy for pairs identification, and forecasting with deep learning seeking better trade entry points.  Similarly \cite{Gat23}, but without the lattermost step.  Many authors appeal to dimension reduction when a large universe of portfolio assets are under consideration; this often amounts to an application of PCA.  (Non-PCA based) Factor models are an alternative with a strong theoretical grounding, and are presented (amongst many other introductory ideas) in the well-known work of  \cite{Vid04}.  Such models are often characterised by a collection of hand-constructed theoretically-justified features called `factors' that to correlate to price or returns time series of the target assets.  An insightful empirical demonstration of both applied PCA and factor models can be found in \cite{Ave10}; also of note, the paper jointly trades a multivariate portfolio of assets -- beyond targeting pairs only.  The approach of \cite{Gui19} is to model returns via a statistical factor model, with an Ornstein-Uhlenbeck process for the error term.  Novelty is introduced at the level of trade execution, formulated via a stochastic control model, though only applied on synthetic data.  Approaches to optimizing mean-reverting multivariate portfolios, subject to either budget or leverage constraints, and influenced by Modern Portfolio Theory, are found in \cite{Zhao18,Zhao19}. An approach to multivariate statistical arbitrage modelling beyond factor models is found in \cite{Stu16}, that presents a framework for both linear and non-linear relationship modelling via vine copulas.

\vspace{0.25 \baselineskip}

\noindent  State space models combine knowledge of the data process with noisy measurement data; a focus of this work is to consider their utility for statistical arbitrage applications.  Hence our eventual contribution -- a filtering approach to multivariate statistical arbitrage, scalable to large portfolios.  Indeed, state space models for finance have been reviewed in \cite{Lau03,Date11}.  The articles include applications to stochastic volatility modelling, and the modelling of the term structures of commodity prices and interest rates.  Applications to pairs trading are relevant, and were introduced by the seminal work of Elliott \cite{Ell05}, who models a latent pair spread, and noisy observations thereof, in a state space setting.  Their results are presented for synthetic data.  The work of \cite{Tri11,Tsa10} extends the research by introducing time-varying parameters to the model, hence improving model flexibility, and testing the theory in realistic online trading settings.  The Elliott model is also generalised somewhat for richer spread dynamics in \cite{dem16}, which designs a novel trading strategy based on model-estimated probabilities of large dislocations mean reverting.  Further generalisations are apparent in the recent article \cite{Zha21} which presents a quasi-Monte Carlo estimation algorithm for the state space framework presented.  This allows for estimating models with richer specifications, designed to express stylized features of financial data, including non-Gaussianity and heteroskedastic variance.

\vspace{0.25 \baselineskip}

\noindent It is natural to incorporate (dynamic) factor models into a state space framework, as in \cite{Han06}, which also models stochastic volatility within the state variable; similarly \cite{Chib06}.  Arguments for dynamic factor modelling for statistical arbitrage are given by \cite{Foc16}, though the approach does not model volatility (beyond a white noise process).  PCA for factor mining, and prediction (of note, at the level of asset prices, rather than returns) is a heavily exploited feature of the trading strategy. 

\vspace{0.25 \baselineskip}

\noindent  The studies related to statistical arbitrage are summarised in Figure \ref{fg:R1} below, and include information on the underlying market data analysed, by asset class, time scale and date range.  `SS' denotes modelling in a state space framework, `Target' is given by `pair' for pair-trading examples, and `MV' for multi-variate portfolios, while `Dim. red.' denotes a dimension reduction technique is a feature of the modelling approach.  

\vspace{0.25 \baselineskip}

\begin{table}[ht]
\small
  \centering

\hskip-0.8cm \begin{tabular}{*{7}c}

  \cmidrule(lr){1-1} \cmidrule(lr){2-4}  \cmidrule(lr){5-7}
  Author (Year) & SS & Target & Dim. red. & Asset class & Date range & Time scale \\  
  \cmidrule(lr){1-1} \cmidrule(lr){2-4}  \cmidrule(lr){5-7}
\cite{Kra17} (2017) &  & pair &   & -  & -  & -\\
\cite{Sar20} (2020) &  & pair &  \ding{51}  & Commodities  & 2009--2019  & intraday -- 5min\\
\cite{Gat23} (2023) &  & MV & \ding{51}  & Equities --  Global  & 2011--2021  & daily\\ 

\cite{Vid04} (2004) &  & pair & \ding{51}  & -  & -  & -\\
\cite{Ave10} (2010) &  & MV & \ding{51}  & Equities -- US  & 1997--2007  & daily\\
\cite{Gui19} (2019) &  & MV & \ding{51}  & -  & -  & -\\ 
\cite{Zhao18} (2018) &  & MV &   & Equities -- US  & 2012--2014  & daily\\
\cite{Zhao19} (2019) &  & MV &   & Equities -- US  & 2010--2014  & daily \\

\cite{Stu16} (2016) &  & MV &   & Equities -- US  & 1992--2015  & daily\\

\cite{Ell05} (2005)& \ding{51} & pair &   & -  & -  & -\\  
\cite{Tri11} (2011) & \ding{51} & pair &   & Equities -- US, Commodities  & 1980--2008  & daily\\
\cite{Tsa10} (2010) & \ding{51} & pair &   & Equities -- US  & 1997--2005  & daily\\
\cite{dem16} (2016)& \ding{51} & pair &   & Equities -- US, Brazil  & 2011--2013  & daily \\
\cite{Zha21} (2021) & \ding{51} & pair &   &  Equities -- US, Taiwan, Hong Kong  & 2012--2019  & daily\\

\cite{Foc16} (2016) & \ding{51} & MV & \ding{51}  & Equities -- US  & 1989--2011  & daily \\
  \cmidrule(lr){1-1} \cmidrule(lr){2-4}  \cmidrule(lr){5-7}
  \end{tabular}

\captionof{figure}{Literature themed around statistical arbitrage, toward applications of state space models for multivariate portfolios.}
\label{fg:R1}
\end{table}

\newpage

\section{A dynamic model of returns and factors}

\emph{State space modelling}.   Consider the case of discrete time increments $k=0,1,...$.  Then a linear Gaussian state space model is given by the time-varying system of equations:
\begin{align}
\bx_{k+1} &= \bA_k \bx_k + \beps^{\bx}_k, \label{eqS} \\
\by_{k} &= \bB_k \bx_k + \beps^{\by}_k. \label{eqM}
\end{align}
Here $\beps^{\bx}_k$ and $\beps^{\by}_k$ model independent zero-mean white-noise processes with respective covariances $\bPsi_k$ and $\bOmega_k$, assumed to be known functions of time.  The matrices $\bA_k$ and $\bB_k$ are also assumed to be known functions of time.  We call equations (\ref{eqS}) and (\ref{eqM}) the `state' and `measurement' equations, respectively.  The state $\bx_k$ is assumed to be a latent random variable, of which $\by_k$ denotes a noisy measurement at time $k$.  Given the system parameters, probabilistic beliefs about the state can be inferred, and these beliefs can be updated given measurement data.

\vspace{0.25 \baselineskip}

\noindent \emph{Kalman filter}. By the classic Kalman filter (KF) algorithm, given an initial state $\bx_0$, and the state and measurement equations above, the filter distribution for $\bx_k | \by_{1:k}$ at time $k>0$ can be shown to be $\thicksim MVN( \bmu_k^f, \bSigma_k^f)$ where
\begin{align}
\bmu_k^f &= \bmu_k^p + \bG_k (\by_k - \bB_k \bmu_k^p ), \\ 
\bSigma_k^f &= (\bI - \bG_k \bB_k) \bSigma_k^p.  
\end{align}
These equations describe the mean and covariance in terms of the `Kalman gain' $\bG$ given by: 
$$\bG_k := \bSigma_k^p \bB_k'(\bB_k \bSigma_k^p \bB_k' + \bOmega_k )^{-1}.$$  

\vspace{0.25 \baselineskip}

\noindent Also, the predictive distribution for $\bx_{k+1} | \by_{1:k}$ at time $k$ can be shown to be $\thicksim MVN( \bmu_k^p, \bSigma_k^p)$ where
\begin{align}
\bmu_k^p &= \bA_k \bmu_k^f, \label{Param1}\\
\bSigma_k^p &= \bA_k \bSigma_k^f \bA_k' + \bPsi_k. \label{Param2}
\end{align}

\vspace{0.25 \baselineskip}

\noindent \emph{Unscented Kalman Filter}.   Generalising equations (\ref{eqS}) and (\ref{eqM}), we consider the non-linear system dynamics
\begin{align}
\bx_{k+1} &= f(\bx_k; \bA_k) + \beps^{\bx}_k, \label{eqSnl} \\
\by_{k} &= g(\bx_k; \bB_k) + \beps^{\by}_k. \label{eqMnl}
\end{align}
Here $f$ and $g$ may be non-linear functions of the state and parameters. An extension of the Kalman filter was developed based on the unscented transform \cite{Jul04UT}, a method for calculating statistics of a random variable that has undergone a non-linear transformation.  The idea behind the algorithm -- called the Unscented Kalman Filter (UKF) -- is to first draw sigma points, which are sample points drawn from the support of an underlying probability distribution.  The sigma points are passed through a non-linear function, so that by the unscented transform the means and covariances of the predictive and filter distributions are estimable; indeed, they are calculable as functions of (weighted) sample means and (cross-)covariances of the transformed points.

\noindent In keeping with the approach of \cite{Wan00}, so for $N$ the state dimension, at initialisation we define vectors of weights $\matrva{w}^c, \matrva{w}^m$ of size $(2N+1)$, and select a set of sigma points denoted $\matrva{\mathcal{X}}$\footnote{There are various suggestions within the literature for how one might choose the weights and sigma points; we keep with the presentation of \cite{Wan00} for our implementation.}.  Each sigma point is transformed on passing through $f$, yielding the set of transformed sigma points $\matrva{\mathcal{Y}} = f(\matrva{\mathcal{X}})$.  Then the predictive distribution for $\bx_{k+1} | \by_{1:k}$ at time $k$ can be shown to be $\thicksim MVN( \bmu_k^p, \bSigma_k^p)$ where
\begin{align}
\bmu_k^p &= \sum_{i=0}^{2N} w^m_i \mathcal{Y}_{i,k}, \label{Param1}\\
\bSigma_k^p &= \sum_{i=0}^{2N}  w^c_i (\mathcal{Y}_{i,k} - \bmu_k^p)(\mathcal{Y}_{i,k} - \bmu_k^p)^T + \bPsi_k. \label{Param2}
\end{align}

\vspace{0.25 \baselineskip}

\noindent Next we calculate $\matrva{\mathcal{Z}} := g(\matrva{\mathcal{Y}})$, and $\hat{\by}_k^- := \sum_{i=0}^{2N} w^m_i \mathcal{Z}_{i,k}$.  Then the filter distribution for $\bx_k | \by_{1:k}$ at time $k$ can be shown to be $\thicksim MVN( \bmu_k^f, \bSigma_k^f)$ where
\begin{align}
\bmu_k^f &= \bmu_k^p + \bG_k (\by_k - \hat{\by}_k^- ),\\
\bSigma_k^f &= \bSigma_k^p - \bG_k \bP_{y,k} \bG_k^T.
\end{align}
In analogy with the Kalman gain defined above, here we have $\bG$ given by: 
\begin{align}
\bP_{y,k} &:= \sum_{i=0}^{2N} w^c_i (\mathcal{Z}_{i,k} -\hat{\by}_k^- ) (\mathcal{Z}_{i,k} -\hat{\by}_k^-)^T  , \\
\bP_{xy,k} &:= \sum_{i=0}^{2N} w^c_i (\mathcal{Y}_{i,k} - \bmu_k^p ) (\mathcal{Z}_{i,k} -\hat{\by}_k^-)^T  , \\
\bG_k &:= \bP_{xy,k} \bP_{y,k}^{-1}.  
\end{align}

\vspace{0.25 \baselineskip}

\noindent  \emph{A state space conditional factor model}. We consider a conditional factor model, as inspired by the Arbitrage Pricing Theory (APT) of Ross \citep{Ross76}.  For a single-period investment setting, we assume the predictive model
\begin{align}
\br_{k+1} = \bX_k \bef_k + \beps_k^r, \qquad \beps_k^r \thicksim N(\bz,\bD_k). \label{eqR}
\end{align}
Here $\br_{k+1}$ is the $n$-dimensional random vector containing the cross-section of excess returns for $n$ equities over the next time increment $[k,k+1]$; $\bX_k$ is a non-random $n \times m$ matrix of observable factor exposures estimated at $k$, for $m$ the number of explanatory asset factors; $\bD_k$ is a covariance matrix for the returns, assumed to be diagonal; and $\bef_k$ is an $m$-dimensional latent factor vector that we estimate at each time step by cross-sectional regression.  

\vspace{0.25 \baselineskip}

\noindent  We write the state space conditional factor model 

\begin{align}
\bef_{k+1} &= g( \bef_k; \btheta^{f}_k) + \beps^{\bef}_k, \label{eqS1} \\
\br_{k+1} &= \bX_k  \bef_k +  \beps^{\br}_k, \label{eqM0} 
\end{align}

\noindent assuming dynamics in the latent factor vector.   The state equation implies the dynamic predictability of $\bef$ via some known function $g$.  Recently, non-linear models for factor evolution have been explored in \cite{Gu21}.  

\vspace{0.25 \baselineskip}

\noindent Of course, a popular modern choice for $g$ might be one of the various machine learning-based time series models.  Such models could depend on a recent history of the state variable, rather than just the current state.  In the case of the factor model above, we can expand the state space to include lags in $f$.  Consider the one dimensional case where
$$
f_k = g(f_{k-1}, \dots, f_{k-M}; \bw) + v_k.
$$
The function $g$, for example, can be estimated by a time series model, with $v_k$ being its aleatoric uncertainty estimate. For this single factor case, per \cite{Wan00}, the state space model can be written:

\newpage

\begin{center}
\makebox[\textwidth]{
  \includegraphics[width = 1.3\textwidth]{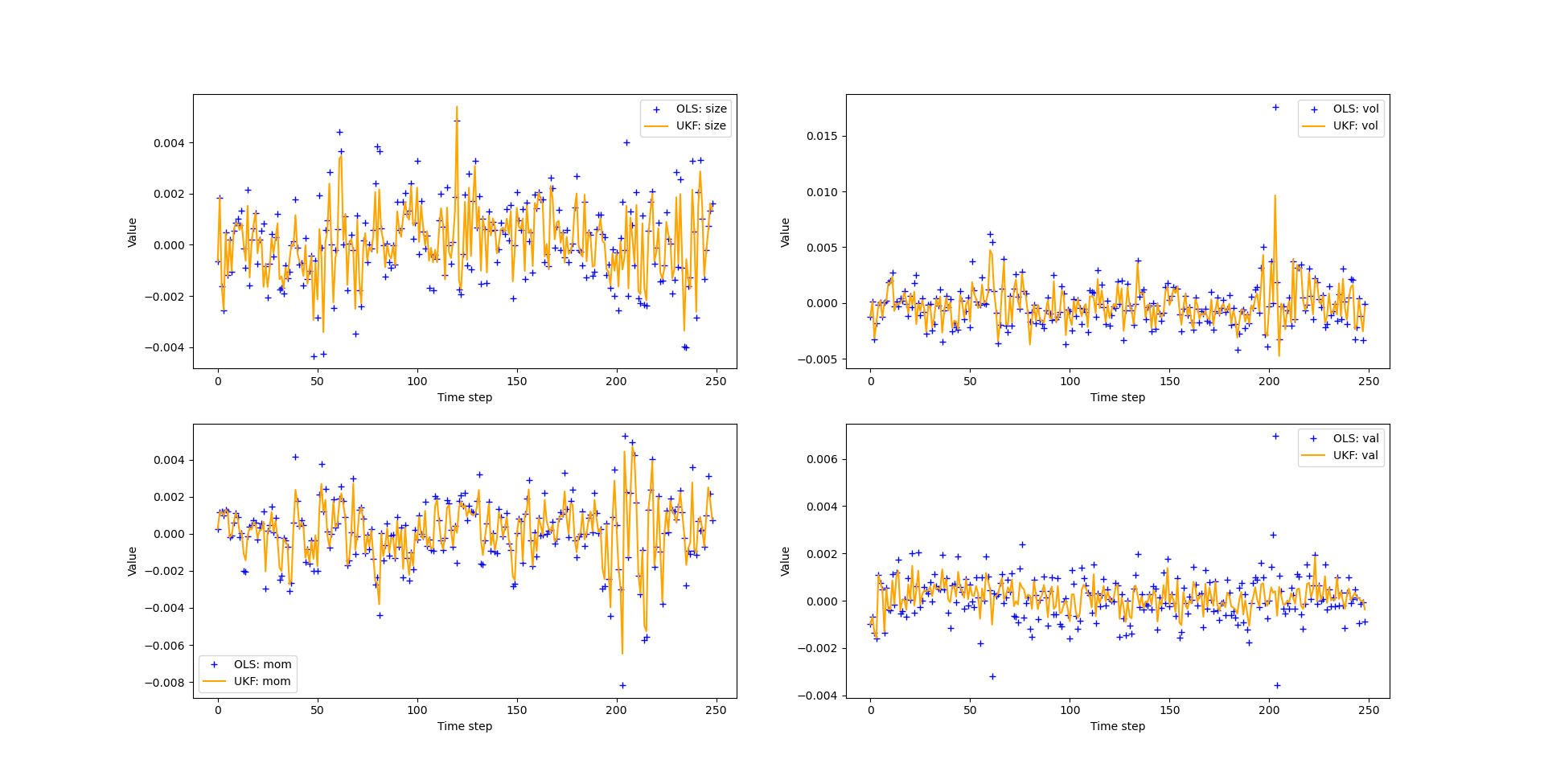}}\par 
  \captionof{figure}{Factor risk premia as estimated by a conditional factor model using (i) ordinary least squares (`OLS'); and (ii) an underlying non-linear state space model (`UKF').  Four factors are displayed -- size, volatility, momentum and value, for the year 1997 (which was arbitrarily chosen).   \label{fg:A}}    
\end{center}

\begin{align*}
\bef_{k+1} &= \bg(\bef_k; \bw) + \bB \beps^{\bef}_k, \\
\begin{bmatrix}
 f_{k+1} \\ 
 f_k \\ 
\vdots \\ 
f_{k-M+2} 
\end{bmatrix} &= \left[\begin{array}{cc}
        &          \\
      \multicolumn{2}{c}{\smash{\raisebox{.5\normalbaselineskip}{$ g(f_k, \dots, f_{k-M+1}; \bw)$}}} \\
      \begin{bmatrix}
 1 & 0 &  \dots & 0 \\ 
 0 & \ddots & 0  & \vdots \\ 
 0 & 0 &  1 & 0
\end{bmatrix} & \begin{bmatrix}
 f_k \\ 
\vdots \\ 
f_{k-M+1} 
\end{bmatrix}
    \end{array}\right] + \begin{bmatrix}
1 \\ 
0 \\ 
\vdots \\ 
0
\end{bmatrix} v_k, \\
r_{k+1} &= \left[\begin{array}{cccc}
1 & 0 & \dots & 0
    \end{array}\right]  f_k + \epsilon^r_k.
\end{align*}

\noindent We demonstrate an application of the model for four latent factors, assumed uncorrelated, showing a sample of model output in Figure \ref{fg:A} above.  This compares with a predictive model for returns as a function of factors estimated by ordinary least squares.  We provide further details of the model specification as part of the following section on experimental methods and strategy.

\section{Experimental methods and strategy}

Our statistical arbitrage scheme has three non-trivial components -- data set construction and feature creation, data modelling, and specification of the trading strategy.  Each of these components can have a significant bearing on performance results.  In Section 3 we introduced the data model, and we present details of the particulars here.  We also address the other 2 components in this section, commenting on the data and the trading strategy.

\vspace{0.25 \baselineskip}

\noindent \emph{Data.}  The data set construction is the same as used in \cite{Spears23}, with the exception that the period is daily rather than bi-monthly.  From that work, recall that the data is sourced from CRSP and IBES databases, with access granted via the Wharton Research Data Service.  The raw data is daily US equity market data for the years 1992-2021 inclusive.  On each trading day, we select the top $N$ greatest stocks across all available US equities sorted by market capitalisation, for $N=2000$.  The data set is filtered for USD denominated common stock only -- there are no closed end funds, REITs, ETFs, unit trusts, depository receipts, warrants etc.  We also collected the S\&P500 (excess) return time series, stock market capitalisations, and earnings per share, adjusted for splits.

\vspace{0.25 \baselineskip}

\noindent In that we experiment with factor models, the data set requires the creation of our 5 chosen risk factors: market beta, size, volatility, momentum and value.  Calculating size is relatively straightforward in CRSP: the calculation is the product of shares outstanding and the daily close price.  Value was determined by analysing IBES data, and could be merged back to the CRSP database based on the  CUSIP identifier.  We note that risk factor construction is often non-trivial.  Momentum was codable using CRSP data and the definitions of \cite{Asn13}; we calculated the daily compounded 12 month returns sans the most recent 1 month of data.  We created the market Beta and volatility factors using the relevant CRSP data and the details of \cite{Kolm17}.  Indeed, over a two year rolling window we estimate a collection of linear functions by regressing each asset's daily excess returns against the S\&P500 excess returns.  The market Beta is the regression gradient, while the volatility factor is set to be the regression mean-square error.

\begin{wrapfigure}{r}{0.4\linewidth}
    \centering
    \includegraphics[width=0.85\linewidth]{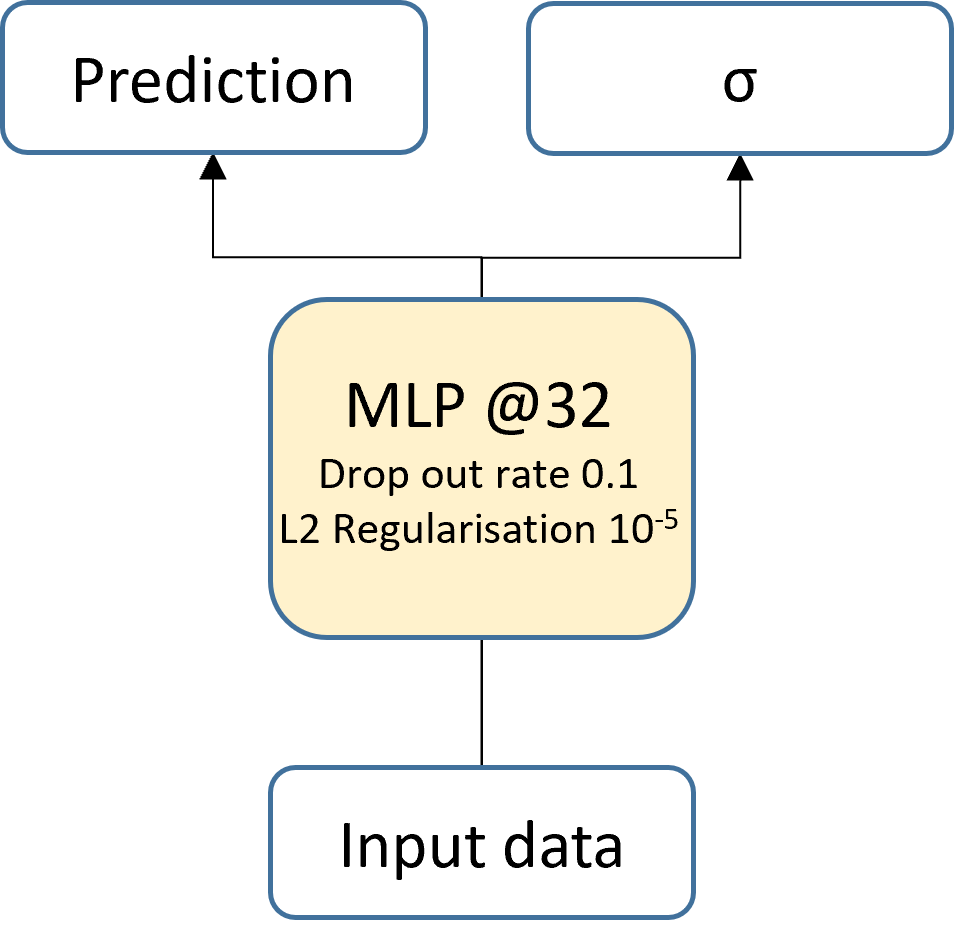}
    \caption{Suggestive neural network architecture for a non-linear latent factor vector state evolution model.  The model outputs state prediction and aleatoric uncertainty estimates given an input historic time series.}
\label{fg:NN}
\end{wrapfigure}

\vspace{0.25 \baselineskip}

\noindent \emph{State space model specification.}  The state space models described in Section 3 require the specification of the state transition function $g$, and the state and measurement covariances, denoted $\bF$ and $\bD$.   In the case of the linear model, we choose $g$ to be the identity function\footnote{The choice of $g$ is in analogy with the state space model underlying the pairs trading application of \cite{Tsa10}.}, and estimate a diagonal state covariance for  $\bef_{k+1} - g( \bef_k; \btheta^{f}_k)$ based on a rolling (lookback) window of size 20.   For the non-linear model, we estimate a multi-layer perceptron model for $g$, that takes the previous history of $f$ on a window of size 10, and that predicts $f$ at the next time-step, and well as the aleatoric uncertainty of $f$, which we set as an estimate of $\bF$.  To be clear, this estimate is a weighted average of the individual uncertainty estimates collected when calculating the state update equation for each sigma point.  For reference, an image of the neural network architecture is given by Figure \ref{fg:NN}.  This model is in the spirit of \cite{Spears21}, and in this simplified setting we implement a multi-layer perceptron with 32 nodes followed by a drop out layer with rate 0.1, and L2 regularisation with parameter fixed at $10^{-5}$ chosen by cross-validation on data from the earliest years.  Finally, for both the linear and the non-linear model, our estimate for $\bD$ is a diagonal covariance gleaned from estimating Eq. (\ref{eqR}) via ordinary least squares regression.

\vspace{0.25 \baselineskip}

\noindent \emph{Benchmark models.} We utilise a benchmark model with the trading strategy mechanics as outlined below, but with the key difference that the spread $S$ is defined by the rolling 10 day sum total returns for each asset against the mean total return over a rolling 10 day window.  This benchmark is discussed in \cite{Lo90}, and is intuitive for considering reversals of returns.  To be clear, this benchmark strategy is with reference to the total return series for each stock only; that is, it makes no reference to any underlying factors.

\begin{center}
\makebox[\textwidth]{
  \includegraphics[width = 1.3\textwidth]{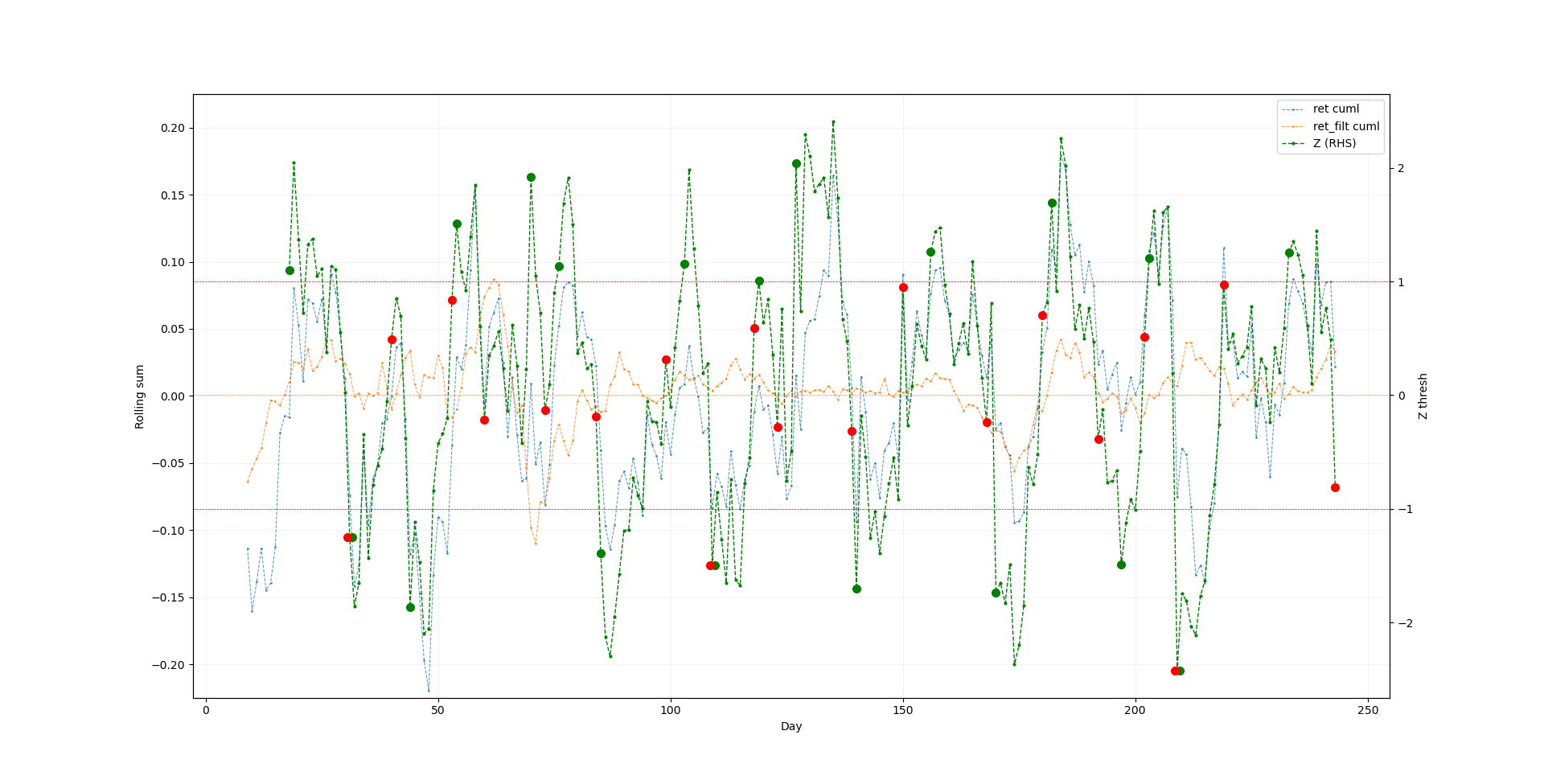}}\par 
  \captionof{figure}{ Trades are entered when the Z score deviation of the difference between returns and filtered returns passes a high (absolute) threshold; these points are indicated by the large green dots.  Trades are closed at the first time the deviation has crossed the origin, as indicated by the large red dots.   \label{fg:A}}  
\end{center}

\vspace{0.25 \baselineskip}

\noindent \emph{Trading strategy.}  The state space model-inspired trading strategy is a long/short strategy, hedged to be beta neutral.  The strategy depends on the spread defined by the rolling sum total returns in excess of the rolling sum filtered total returns -- as estimated using the models of Section 3 -- for each asset.  That is, for each time $k$, $r_k$, $X_k$ and $X_{k-1}$ are observed, so the filter estimate for $f_{k-1}$,  denoted by $f^{(\text{filt})}_{k-1},$ is calculable.  Hence we can calculate $r^{(\text{filt})}_k = X_k f^{(\text{filt})}_{k-1}$. The spread, or (sum) error deviation, $\sum (r_k - r^{(\text{filt})}_k)$ is the target for our trading strategy, and we implicitly assume that it is mean-reverting.  

\vspace{0.25 \baselineskip}

\noindent Being arbitrary, the rolling window size for calculating the spread, denoted $WS$, is a target parameter for performance sensitivity analysis.  Given a trading threshold $L$ ($S$), a trade is entered long (short) the first time the Z score of spread divergences is below (above) the negative (positive) values of $L$ ($S$). The trade is exited the first time the Z score reverts to, or through, the level $0$, with trades `closed out'  on the last trading day of the year otherwise.  The quantity $L$ ($S$) should represent a sufficiently large deviation such that expected excess returns are not dominated by transaction costs.  All position sizes long or short are set to a constant notional value.  To assist intuition, the mechanics of the strategy are outlined diagrammatically in Figure \ref{fg:A} above, for the choice $L/S = 1.0/1.0$, for a target asset over the course of a year.

\vspace{0.25 \baselineskip}

\noindent There are two other minor assumptions to the trading strategy that we state for completeness.  Firstly, when entering a trade we assume that we can execute at the market close price.  Secondly, at time $t$, if our database contains no close price for a stock at $t+1$, we assume that time $t$ is the last trading day for the stock, and that this is known to market.  Hence, we close any existing position in the stock at the time $t$ close price, and do not enter a new trade at $t$ in the event that the trade entry criteria is met.

\vspace{0.25 \baselineskip}

\noindent \emph{Transaction costs.}  In our primary experiments transaction costs are set to 5 basis points (bps) per trade entry and per trade exit, making 10 bps for a round-trip trade in a single underlying (be it a single stock or index).  This is consistent with the works of  \cite{Stu16,Ave10} and chosen for ease of comparison; these works are important benchmarks for multivariate statistical arbitrage in US equities.  Further, we assume that we are able to borrow to short at all times, and that the transaction costs subsume any cost of borrowing.  This deviates to some extent from reality, where not all stocks can be shorted all the time, and where some stocks incur a `hard to borrow' fee, particularly when in high demand.  Unless specified otherwise, all returns discussed for stocks and indices are total returns, and short positions pay the risk-free rate assuming funding is received at the time of shorting, and able to be invested in the riskless asset.  We test the robustness of our strategies to alternative levels of transaction costs; this is discussed further in the Results section below.

\vspace{0.25 \baselineskip}

\noindent \emph{Hedging market risk.}  We hedge our trading strategy to be beta neutral.   Our simple approach is to calculate the market beta of our portfolio of long and short positions each trade period, and to maintain a position of dollar value $P$ in the S\&P500 index, where 
$$
P = - \beta_{port}  \Pi,  
$$
thus providing the hedge.  The term $\Pi$ denotes the market dollar value of the portfolio to be hedged (and is defined further below).  A positive value of $P$ indicates a net long postion in the index as the hedge, while a negative value indicates a short position.  The term $ \beta_{port}$ denotes the weighted-average beta of the portfolio given by
$$
\beta_{port} =  \frac{ \bbeta \bD^T  }{ \Pi  }.  
$$
The term $\bD$ is an $N$-vector of position sizes whose $i^{th}$ entry, consistent with the trading strategy notes above, takes the value $1$,$-1$ or $0$ for a long, short or nil position in stock $i$.  The term $\bbeta$ is an $N$-vector of corresponding stock beta estimates.  

\vspace{0.25 \baselineskip}

\noindent Implicit in the hedging strategy is the assumption that we can trade long or short positions in the S\&P 500 index and earn or pay the total return of the index each period.  In reality, the investor would take a position in the futures contract written on the index, which has a myriad of practical implications.  Regardless, we do not consider these further, and so, for example, we ignore the spot / futures basis that could induce additional slippage.  

\vspace{0.25 \baselineskip}

\noindent \emph{Strategy returns.} We assume a 100\% leverage portfolio so that for each dollar invested in the portfolio, a position of up to a dollar's worth long and / or a dollar's worth short can be entered.  Therefore, the market cost for $l$ longs and $s$ shorts is $\Pi = \max(l,s)$.  We calculate (excess, unhedged) portfolio returns for a time period $[k, k+dt]$ thus:  
$$
\Delta \Pi_k / \Pi_k = \frac{\sum_i (R^l_{i,k} - r_k dt) + \sum_j (R^s_{j,k} - r_k dt) }{\Pi_k} + \frac{r_k dt S_k}{\Pi_k} -   \frac{TC \cdot \bo \cdot ( \Delta \bD_k )^{|\cdot|}}{\Pi_k}.
$$

\noindent Here $R^l_{i,k}$ ($R^s_{j,k}$) denotes the return over the period for a long (short) position  held,  $r_k$ denotes the (annualised) risk-free rate of return,  $TC$ denotes the transaction cost, and $\bX^{|\cdot|}$ defines an operator returning a vector of elementwise absolute values of the input $\bX$.  Hence the terms on the right-hand side of the summation account for, respectively, the excess return of the long and short positions, the risk free rate earned on the proceeds of the short sales deposited in a money account, and the transaction costs for changes to the portfolio holdings\footnote{The calculation is similar to that found in \cite{Ave10}, and is equal when $\Pi = s$.  Otherwise, our return is slightly more conservative for the case $\Pi = l$, though this has little impact on the final value calculated.}.

\newpage 

\section{Results}

In this section, we present performance statistics for the state space factor models of Section 3, relative to our benchmarks and the long-only market portfolio.  For readers inclined, it is possible to replicate the results given the descriptions of the data and methodology as outlined in Section 4.  

\vspace{0.25 \baselineskip}

\noindent We find that the multivariate statistical arbitrage strategy has shown reasonable performance through time, which is consistent with the expectations for a mid-frequency strategy with reasonable capacity.  Performance is best during the first half of our 29 year sample, and we conjecture that the impressive returns have diminished due to increased market competition and efficiency through time.  Also, we find sufficient differences between strategy performance that supports the use of a state space conditional factor model for multi-variate statistical arbitrage, instead of other less complex - albeit relatively powerful - models.  Despite all strategies showing relative underperformance in the most recent years, practitioners may find latitude for improvement by further modifying or optimizing components of the strategy; we briefly describe some preliminary ideas for research direction at the end of this section.

\subsection{Strategy performance}

\noindent \emph{The earlier years: 1993--2007.}  Given the data set and assumed fair value models, per the methods described in Section 4 there is an implied sensitivity of any estimated performance statistics to the key assumptions underlying the trading strategy.  Not the least, this includes the setting of parameters denoting (i) the trade entry threshold Z score; (ii) the size of the returns window; and (iii) the level of transaction costs.  We set the parameters of (i) and (ii) based on strategy performance over the `earlier years' of 1993-2007.  We assume a transaction cost level of 5 bps for each trade in all results that follow, unless otherwise specified.  Choosing the parameters in this way also allows for a reasonable performance comparison between our strategy relative to results given by other well known academic works.  Finally, the parameter choices are maintained when we assess strategy aggregate performance, and performance conditioned on later years only.

\vspace{0.25 \baselineskip}

\noindent For various combinations of Z score-based entry thresholds for the long and short legs (denoted L and S), and various window sizes (denoted WS), we show average Sharpe ratio and maximum drawdown statistics in Figure  \ref{fg:I}, on the next page.  Drawdowns are calculated on a 252 day rolling window.  We find utility in solving for the trade entry parameters for the long and short legs separately, making no assumption that the same cut off should be optimal for both.  We test the entry level to balance entering a sufficient number of trades, but not entering too early (or late) such that the trade is not held too long (or missed entirely).  As expected, we find firstly that as the threshold decreases, the number of trades increases to the extent that small thresholds have bad economics owing to large turnover and cumulative transaction costs.  Window size is similarly tested, for sizes of 1, 5 and 10, corresponding to daily, weekly and fortnightly rolling windows.

\vspace{0.25 \baselineskip}

\noindent  Per Figure \ref{fg:I}, we notice two clear clusters of data, whereby (i) one exhibits the highest average Sharpe ratio, but also exhibits amongst the worst for maximum drawdowns; and (ii) the other is a cluster with good average Sharpe ratio, but amongst the strategies with the better of the maximum drawdowns.  It is reasonable to suggest that different investors might have different preferences for either cluster, so we extend our study for recent years with respect to them.  The first cluster is characterised by models with a window size of 5, and long/short entry thresholds in \{0.5,1.0,1.5,2.0 : $L<S$\}.  The second cluster is the same except that the window size is 10.  For both clusters, the state space models tend to dominate both of the benchmark models with respect to both statistics of interest.  However, the performance between the state space models is less noticeable, though is such that the KF tends to have the higher Sharpe, for worse drawdowns, relative to the UKF, and for all choices of cutoff.  Regardless of an investors preference, these differences are relatively small for this experiment.

\newpage

\noindent
\begin{minipage}{\textwidth}
\begin{center}
\makebox[\textwidth]{
  \includegraphics[height = 1.0\textheight, width = 1.1\textwidth]{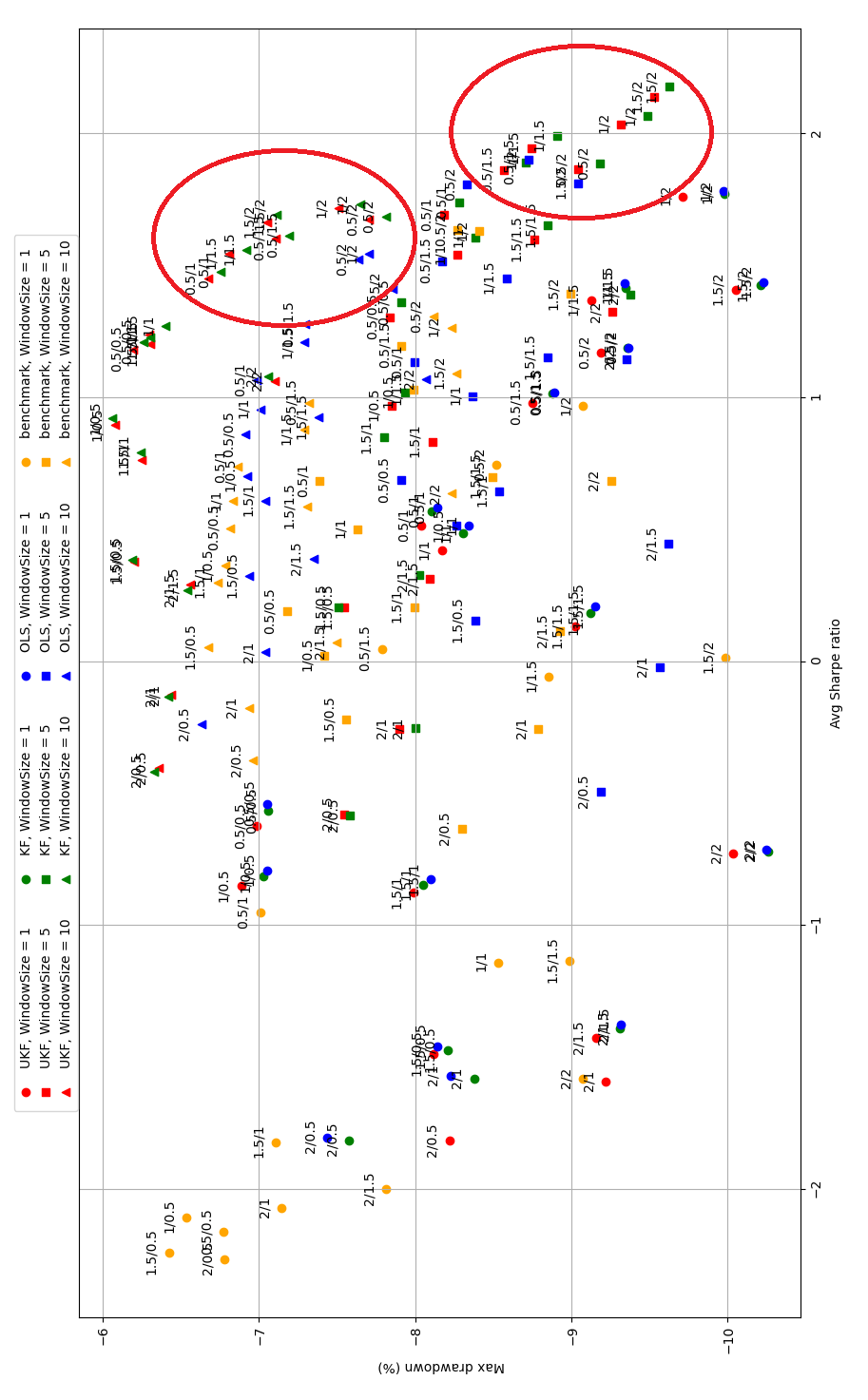}}\par 
  \captionof{figure}{Average Sharpe ratio over the 15 years from 1993-2007 versus maximum drawdown for a collection of models underlying the trading strategy.  
\label{fg:I}}
\end{center}
\end{minipage}

\newpage

\begin{table}[ht]
\small
\centering
\captionof{figure}{Tabulated average annual Sharpe ratios by date range, for our conditional factor model (hedged UKF,  L/S = 1.5/2.0, WS=5) and two leading external methods.  Note I: All methods assume transaction costs of 5bps per trade.  Note II: * results mean `excluding 1992 due to lack of data'. \\ }
\label{fg:R2}
\begin{tabular}{*{6}c}
  \cmidrule(lr){1-6} 
  Strategy & Date range & \multicolumn{2}{c}{ Avg ann Sharpe} & TR SPX &  Notes (Models for cited work) \\ 
    & &\cite{Ave10, Stu16} & Ours  & &  \\
  \cmidrule(lr){1-6} 
Ours  & 1993-2021  & -  & \bf{1.29} & 0.87 & - \\
  \cmidrule(lr){2-6} 
 \cite{Ave10}  & 1997-2007  & 1.44  & \bf{1.61} & 0.47 &  PCA based \\
  & 2003-2007  & 0.90  & \bf{1.01} & 0.78 & PCA based \\
  & 1997-2007  & 1.10  & \bf{1.61} & 0.47 & ETF based \\
  & 2003-2007  & \bf{1.51}  & 1.01 & 0.78 & ETF based --  incorporating additional volume feature \\
  \cmidrule(lr){2-6} 
\cite{Stu16}& 1992-2015  & 0.75-1.12  & \bf{1.67}* & 0.73*& Vine based. Range across 4 alt. trade selection methods.\\
  \cmidrule(lr){1-6} 
\end{tabular}
\end{table}

\vspace{0.25 \baselineskip}

\noindent One also notices from Figure \ref{fg:I} that the strategies that have a lower entry threshold for the longs, and relatively larger entry threshold for the shorts, tend to exhibit the better performance statistics.  It is the case that the longs outperform the shorts for this data sample\footnote{\noindent Recently, other authors have analysed long-short equity factor portfolios, albeit in an alternate strategy setting, and made the same outperformance observation \cite{Blitz20}.}, with the shorts also more likely to have a negative year of performance.  The effect of increasing an entry threshold is to reduce the number of trades entered on that long or short leg; this reduces relatively positive or negative returns, hence the observation.  We leave the question of why the longs may more often outperform the shorts as future work, and include no assumption as to what leg may outperform for any period within our backtest; this could possibly be a source of improvement of the strategy alpha.

\vspace{0.25 \baselineskip}

\noindent Our results can be reasonably compared with other well-known results from the literature.  Our approach has a similar data set to the others (daily US equities data, albeit with different universe sizes), but most notably different models for fair-value -- in \cite{Ave10} factors are modelled based on Principal Component Analysis, with an additional ETF-based model, and in \cite{Stu16} asset relationships are modelled with Vine copulas.  Our trading strategy is similar to \cite{Ave10}, though we do not refine a trade exit threshold; we have less in common with \cite{Stu16}, who refine their trading rules with many more steps relative to our approach.  Importantly, across all papers, the transaction cost assumption is equal.  For the papers cited, it is easiest to compare average annual Sharpe ratios over the date ranges given in the cited paper.  We tabulate the results of this comparison in  Figure \ref{fg:R2} above, and also include the total (excess) return of the S\&P 500 over the same period.  Our results are similar in magnitude to \cite{Ave10}, and we conjecture that daily US equity mean reversion could be captured by strategies using a variety of underlying fair value models over the considered date range.  Superficially, there appears to be some outperformance in our approach, though this claim would benefit from comparison given outcomes of a controlled experiment.  On the other hand, though the average Sharpe ratios appear similar, our methods do show sufficient variation in Sharpe ratio year-on-year.  For example, the PCA based method shows severe performance degradation, and negative Sharpes, in later years.  On the other hand, our approach had no negative years between 1997 and 2007.  The outperformance of our approach is more striking against \cite{Stu16}, and with many degrees of freedom in the formulation of the copula-based strategy it is hard to conjecture exactly why this may be the case.

\vspace{0.25 \baselineskip}

\noindent \emph{Aggregate performance.}  In Figure \ref{fg:II} below, we depict the annual Sharpe ratio distribution over 29 years by way of box and whisker plots, for the 4 underlying models.  Further, we show the performance across 4 sets of parameter choices, which include window sizes of 5 and 10, and long/short entry thresholds of 0.5/2.0 and 1.5/2.0.  We also show a plot for the Sharpe ratio of total (excess) return of the S\&P 500.  

\newpage

\begin{center}
\makebox[\textwidth]{
  \includegraphics[width = 1.35\textwidth]{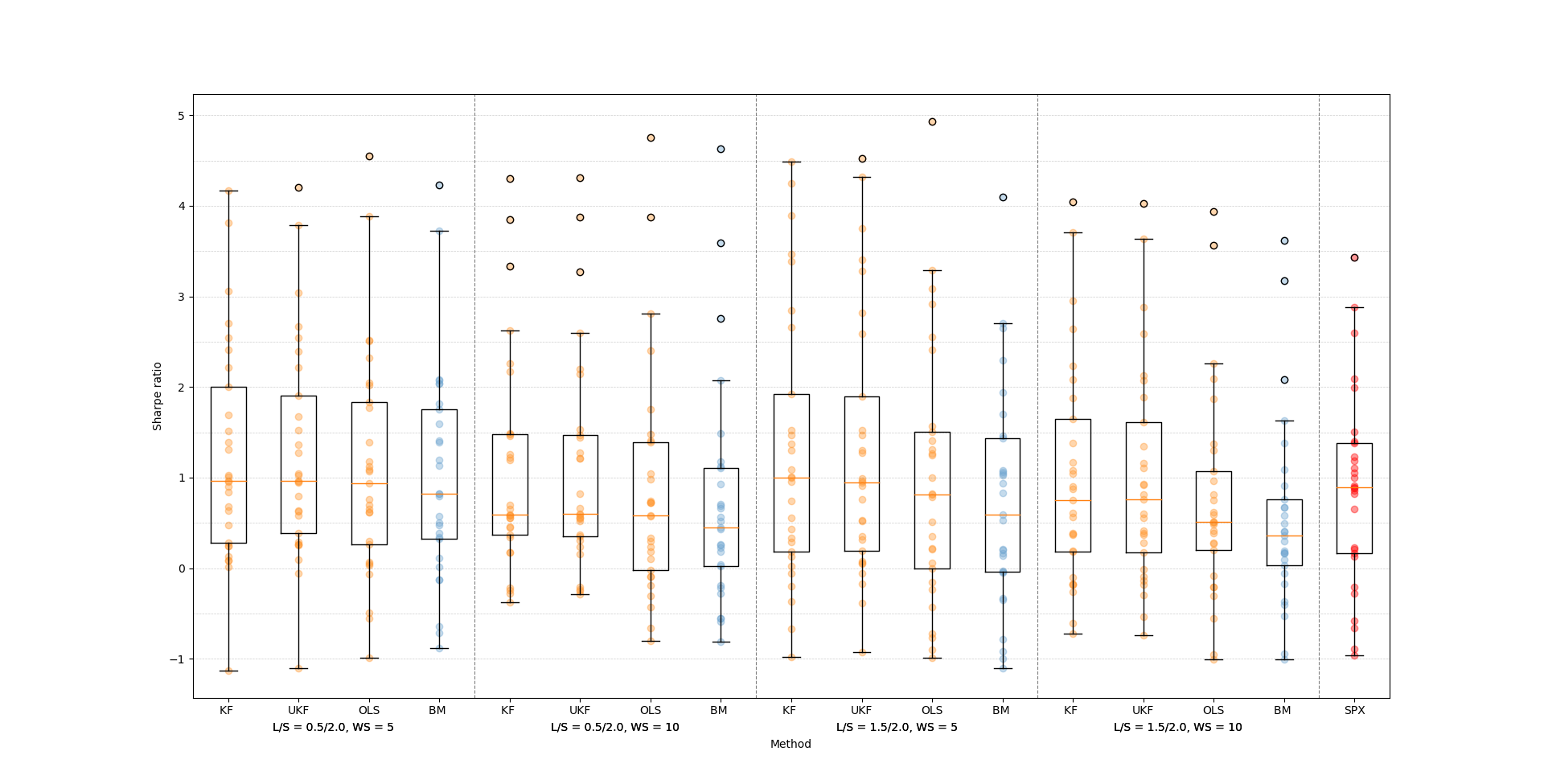}}\par 
  \captionof{figure}{Sharpe ratio distribution over 29 years for 4 models underlying the trading strategy outlined in Section 3, grouped by long/short entry Z score criteria, and lookback window size for cumulative returns.  The performance of the S\&P 500 with respect to total (excess) return is also included for reference.  
\label{fg:II}}
\end{center}

\begin{wraptable}{r}{0.5\textwidth}
  \centering
 \begin{tabular}{*{6}c}
  \cmidrule(lr){1-6} 
   & &\multicolumn{4}{c}{ Transaction costs (bps)}  \\ 
   & & 0 & 5 & 10 & 15  \\ 
  \cmidrule(lr){1-6} 
  & 25th &  1.11  & \cellcolor{lemoncolor}  0.39  &  -0.08 & -0.76   \\
WS 5 & 50th  & 1.39   & \cellcolor{lemoncolor} 0.99  & 0.57  & -0.09   \\
  & 75th &  2.75  & \cellcolor{lemoncolor} 1.91  &  1.14 & 0.70    \\
  \cmidrule(lr){2-6} 
  & 25th &  0.80  &\cellcolor{lemoncolor} 0.35  &  -0.10 & -0.66  \\
WS 10 & 50th  & 1.09   &\cellcolor{lemoncolor} 0.59  & 0.26  & -0.17  \\
  & 75th &  1.90  & \cellcolor{lemoncolor}1.47  &  1.05 & 0.83    \\
  \cmidrule(lr){1-6} 
  \end{tabular}
\captionof{figure}{Sharpe ratio percentiles by transaction cost and window size levels for UKF 0.5/2.0.}
\label{fg:R420}
\end{wraptable}


\noindent Modelwise, the KF and the UKF outperform relative to OLS and BM across all but one groupwise comparisons of 1st, 2nd and 3rd quantiles, and averages, of annual Sharpe ratios.  On the other hand, any UKF under/outperformance over the KF is not apparent, so that the additional computational complexity of the UKF is not justified for this particular application. If stronger non-linear relationships between variables were observed, or expected, given additional training data, the UKF may still find use.  The OLS approach induces good results, with the (U)KF appearing to be a worthwhile refinement, that is, the introduction of uncertainty quantification has a benefit.  The nature of the outperformance appears to vary depending on the parameter settings of the trading strategy.

\vspace{0.25 \baselineskip}

\noindent In Figure \ref{fg:R420}, we show the results of an exploration of robustness of our strategy to the transaction cost assumption.  As mentioned, we implemented our strategies assuming a 5 bps (10 bps round-trip) transaction cost for each asset traded.  We show Sharpe ratio percentile levels for transaction costs of 0,5,10 and 15 bps for the UKF 0.5/2.0 across window sizes.  The yellow highlighted data are for the 5 bps transaction cost level.  Clearly this is a sweet spot -- and while the quoted performance can sustain a slightly higher 

\begin{center}
\makebox[\textwidth]{
  \includegraphics[width = 1.2\textwidth]{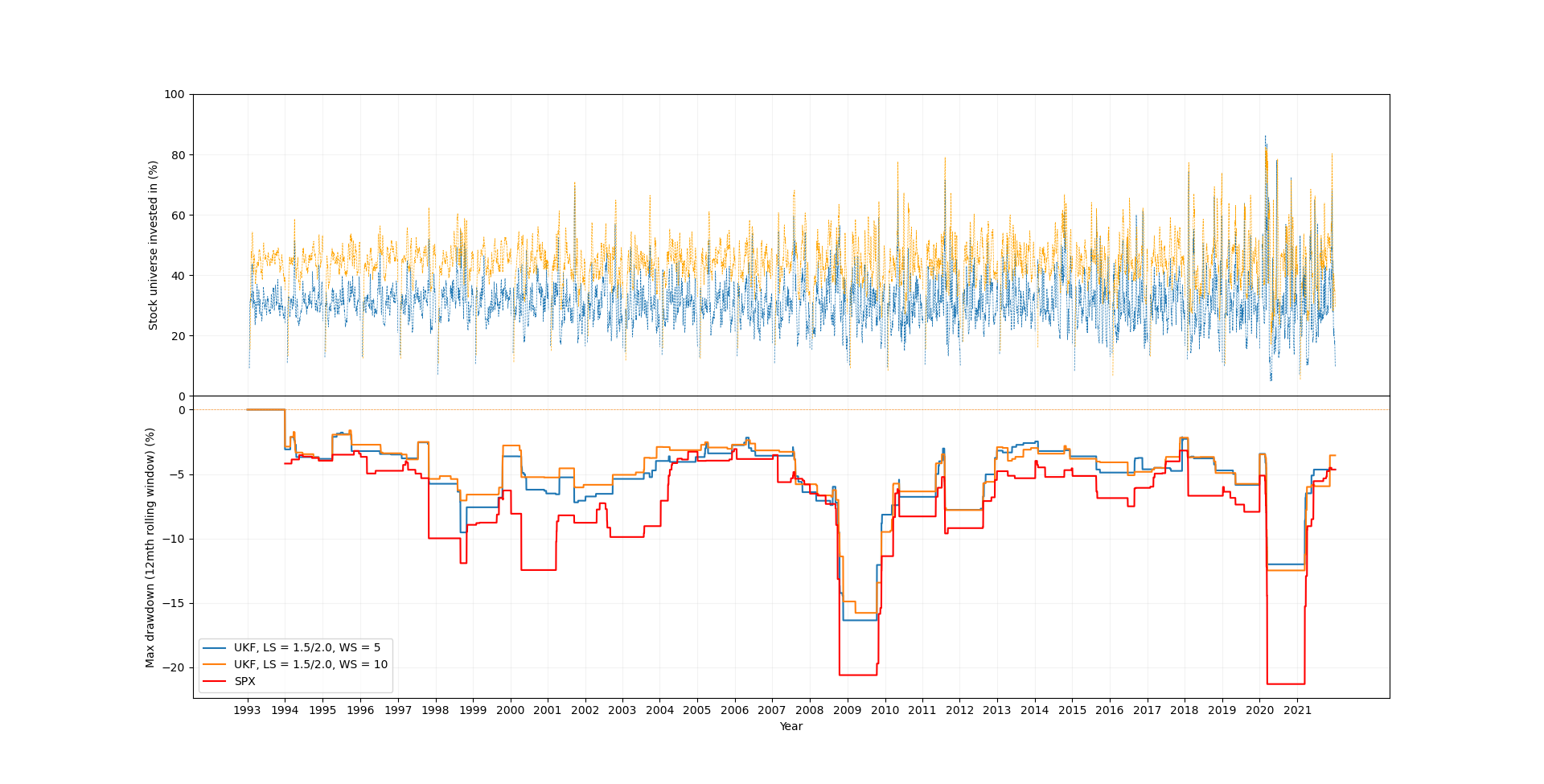}}\par 
  \captionof{figure}{Top: Percentage of the stock universe (N=2000) invested in, long or short, at a given time.   Bottom: Strategy rolling window (252 day) drawdown versus the long only market portfolio; transaction cost = 5bps.  \label{fg:III}}
\end{center}

\noindent transaction cost assumption, there is an obvious performance decay such that at 15 bps of transaction costs, the median annual Sharpe ratio over the sample realises negative.  Finally, we should mention that the cost of short selling may have been significantly more than estimated in this work, if the capacity to borrow to sell was even available, particularly so for the earlier years of study.  On the other hand, practitioners may have strategies for cheapening transaction costs when trading.  Given these latter points, and other cited academic studies, we think our primary transaction cost assumption is reasonable.  

\vspace{0.25 \baselineskip}

\noindent The bottom panel of Figure \ref{fg:III} shows rolling 252 day drawdown for 2 UKF based strategies differing on window size.  In the years up till 2007 the strategies largely avoid the worst of the drawdowns realised by the market portfolio, which has drawdowns touching -10\% or worse on 3 occasions.  In these cases, our strategies realise drawdowns between -7 and -5 \%, with the exception of the strategy with window size of 5 that realises a -9.5\% drawdown on one of the three occasions.  Post-2007, the two severe market drawdowns of 2008 and 2020, both being worse than -20\%,  are matched by strategy drawdowns about -16 and - 12\%, respectively.  The first of these drawdowns, within the years of the Great Financial Crisis, was to the chagrin of many real-world strategies, such that we find it less of a concern that our strategy did not register greater outperformance during this period.  For intuition, we also depict the proportion of assets under consideration that are invested in for all times, in the top panel of Figure \ref{fg:III}.  We notice the obvious empirical fact that with a larger window size, more assets are held.  There is an annual seasonality apparent, whereby at the end of each year positions are closed out, and at the start of the year there are spikes for new trades re-entered, as expceted by the definition of our trading strategy.

\vspace{0.25 \baselineskip}

\noindent It is reasonable to assume that the beta neutral statistical arbitrage strategy has zero or low (absolute) correlation of returns to the long only market portfolio.  Hence, in theory, a blended investment taking a position in the strategy and the long only portfolio could show an outperformance (with respect to Sharpe ratio) 

\begin{center}
\makebox[\textwidth]{
  \includegraphics[width = 1.2\textwidth]{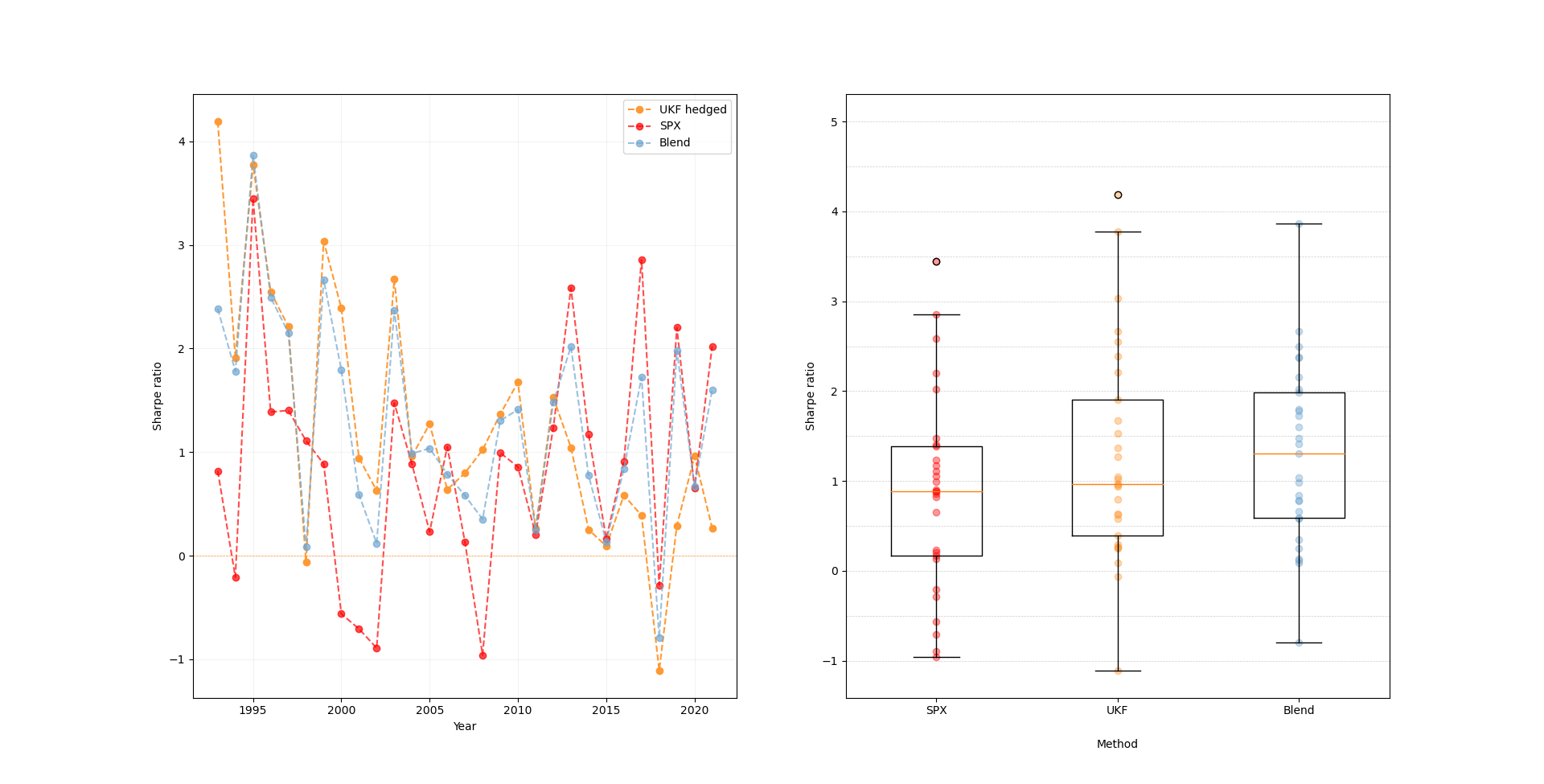}}\par 
  \captionof{figure}{Sharpe ratio distribution over 29 years for a UKF model-based trading strategy as outlined in Section 3, the S\&P 500 with respect to total (excess) return, and a portfolio of a weighted combination of the two.  \label{fg:IV}}
\end{center}


\noindent relative to both \cite{Bru97}.  We implement such a blended strategy by weighting the statistical arbitrage strategy and the long only portfolio as
$$
w_i = \frac{S_i}{S_L + S_{LS}}, \qquad S_i := \frac{\mu_i}{ \sigma^2_i}, \qquad i \in \{ \text{L, LS} \},
$$
where the weights $w$ depend on the mean $\mu$ and variance $\sigma^2$ of the long portfolio $L$ or long-short portfolio $LS$ returns, optimizing for maximum Sharpe ratio \cite{Bru97, Spears21}.  The true values of the mean and variance parameters are unknown, and for demonstration purposes we take the straightforward approach of setting the means and variances equal in the first year, and equal to their historical values on a rolling window of size (up to) 10 otherwise.  We find that the results are not particularly sensitive to (reasonable) values of the choice of rolling window size.  Images summarising the Sharpe ratio performance are given in Figure \ref{fg:IV}.  Of note, given the right-hand side chart, we notice considerable performance improvement with respect to the blended strategy for the first, second and third quartiles, supporting the theoretical result, and offering a practical hint for deploying such a strategy.

\vspace{0.25 \baselineskip}

\noindent \emph{Recent performance: 2008--2022.} We notice from the left-hand side image of Figure \ref{fg:IV} that the strategy has not performed so well in the most recent years, particularly for the years 2015-2019.  We conjecture that this could be due to a lower volatility market regime during this time, increased sophistication of market participants and with it competition for alpha / eroded edge, or otherwise due to limitations derived from some underlying assumption of our strategy.  Overcoming these challenges is likely of interest to the motivated academic or practitioner, and we offer some avenues for future work in the following conclusion.

\newpage

\section{Conclusion}

We study multivariate statistical arbitrage under a conditional factor model augmented in a state space framework for US equities trading.  Our study shows the modelling approach can yield compelling performance statistics over a 29 year period, an an absolute basis and relative to a reasonable benchmark, and relative to a model estimating returns as a function of factors via ordinary least squares.  However, for our experimental set up, we did not find that the non-linear state space model was justified relative to the linear case; though it was no worse, any outperformance benefit does not contrast well with the increased modelling complexity introduced.  When we compare our results to existing attempts in the literature, we find that our approach compares well.  We also show empirical evidence that blending our investment capital between the long-short strategy and the long-only market portfolio as a function of each strategies estimated mean and variance yields a strategy with improved annual Sharpe ratio, consistent with theory.  All results are with respect to reasonable level of transaction costs.

\vspace{0.25 \baselineskip}

\noindent There is the scope for academic and practioner work on the strategy presented in this paper, particularly with respect to exploring the potential for improvement on the period of strategy underperformance in recent years.   We conclude with some final remarks:

\begin{itemize}
\item  Regarding the data underlying the quantitative strategy, the addition of alternative data beyond factor data and returns time series may improve the predictive power of the model.  With respect to the dataset we constructed, and by the recent work of \cite{How23}, it is the case that estimating submodels over clusters of data -- for example, stocks grouped by sector -- could lead to significant performance improvement.  

\item  The recent work of  \cite{Gu21, And23} consider alternative specifications of dynamic factor models for investing, including those with non-linearities at the level of the measurement equation.  Such expressions could be handled by our UKF-based framework, and present as a reasonable target for improving performance.  

\item Recall our brief discussion on related literature in Section 2, whereby almost all work cited made use of data on a daily time scale.  The literature would be bolstered by analysis completed on alternative time frames, particularly on intraday time scales.  This is arguably harder, not the least in terms of data set curation.  On the other hand, the recent work of \cite{Ale23} provides evidence of intraday return predictability as a function of factors using machine learning; this could support refining reactive entry and exit rules within the trading strategy.

\item Modelling techniques subsuming regime filters could support our statistical arbitrage trading strategy, with recent evidence to this end for a pairs trading application given in \cite{Ell18}.

\item Finally, and requiring more expert / domain knowledge, it would be interesting to understand trades entered that are directly mappable to a market phenomena, such as a stock going ex-dividend, or being influenced by index rebalancing. This is interesting in the context of determining whether the quantitative strategy's profitable trades are explainable in terms of a well-known market effect, or else is supportive of the underlying factor-based theory.

\end{itemize}

\section*{Acknowledgements}

The authors would like to thank the Oxford-Man Institute of Quantitative Finance for its generous support.  SR would like to further thank the Royal Academy of Engineering.

\newpage
\setlength{\bibsep}{1pt}
\small\bibliography{bibs}{}
\bibliographystyle{unsrtnat}

\end{document}